\documentclass[fleqn,10pt]{wlscirep}
\usepackage[utf8]{inputenc}
\usepackage[T1]{fontenc}
\usepackage{hyperref}
\usepackage[final]{pdfpages}
\makeatletter
\AtBeginDocument{\let\LS@rot\@undefined}
\makeatother
\hypersetup{
    colorlinks = false,
    linkbordercolor = {white},
}

\title{Structural roles and gender disparities in corruption networks}

\author[1]{\normalsize Arthur A. B. Pessa} 
\author[1]{\normalsize Alvaro F. Martins} 
\author[1]{\normalsize M\^onica V. Prates} 

\author[2]{\normalsize Sebastian~Gon\c{c}alves} 

\author[3]{\normalsize Cristina~Masoller} 

\author[4,5,6,7,8,*]{\normalsize Matja{\v z}~Perc} 

\author[1,*]{\normalsize Haroldo V. Ribeiro} 

\affil[1]{\footnotesize Departamento de F\'isica, Universidade Estadual de Maring\'a, Maring\'a, PR 87020-900, Brazil}

\affil[2]{\footnotesize Instituto de F\'isica, Universidade Federal do Rio Grande do Sul -- Porto Alegre, RS 91501-970, Brazil}

\affil[3]{\footnotesize Departament de F\'isica, Universitat Polit\'ecnica de Catalunya, Rambla St. Nebridi 22, Terrassa, 08222, Barcelona, Spain}

\affil[4]{\footnotesize Faculty of Natural Sciences and Mathematics, University of Maribor, Koro{\v s}ka cesta 160, 2000 Maribor, Slovenia}
\affil[5]{\footnotesize Community Healthcare Center Dr. Adolf Drolc Maribor, Ulica talcev 9, 2000 Maribor, Slovenia}
\affil[6]{\footnotesize Department of Physics, Kyung Hee University, 26 Kyungheedae-ro, Dongdaemun-gu, Seoul 02447, Republic of Korea}
\affil[7]{\footnotesize Complexity Science Hub, Metternichgasse 8, 1030 Vienna, Austria}
\affil[8]{\footnotesize University College, Korea University, 145 Anam-ro, Seongbuk-gu, Seoul 02841, Republic of Korea}

\affil[*]{\footnotesize email: matjaz.perc@gmail.com, hvr@dfi.uem.br}

\begin{abstract}
Criminal activities are predominantly due to males, with females exhibiting a significantly lower involvement, especially in serious offenses. This pattern extends to organized crime, where females are often perceived as less tolerant to illegal practices. However, the roles of males and females within corruption networks are less understood. Here, we analyze data from political scandals in Brazil and Spain to shed light on gender differences in corruption networks. Our findings reveal that females constitute 10\% and 20\% of all agents in the Brazilian and Spanish corruption networks, respectively, with these proportions remaining stable over time and across different scandal sizes. Despite this disparity in representation, centrality measures are comparable between genders, except among highly central individuals, for which males are further overrepresented. Additionally, gender has no significant impact on network resilience, whether through random dismantling or targeted attacks on the largest component. Males are more likely to be involved in multiple scandals than females, and scandals predominantly involving females are rare, though these differences are explained by a null network model in which gender is randomly assigned while maintaining gender proportions. Our results further reveal that the underrepresentation of females partially explains gender homophily in network associations, although in the Spanish network, male-to-male connections exceed expectations derived from a null model.
\end{abstract}

\begin{document}
\rfoot{\small\sffamily\bfseries\thepage/13}%

\flushbottom
\maketitle
\thispagestyle{empty}

\section*{Introduction}\label{sec:introduction}

Criminology literature consistently highlights substantial disparities in offending rates between males and females, with males being responsible for the majority of crimes~\cite{bennett2005explaining, kruttschnitt2013gender}. This pattern extends to corruption-related offenses, where experimental research suggests that females typically demonstrate lower tolerance for dishonesty and exhibit more pro-social behavior~\cite{swamy2001gender, dollar2001arewomen, rivas2013anexperiment}. Controlled experiments further indicate that, on average, females offer smaller bribes than males, and interactions exclusively between females tend to exhibit higher levels of honesty compared to those involving at least one male~\cite{rivas2013anexperiment, chaudhuri2012gender}. Additionally, studies report positive correlations between female participation in government and reduced perceptions of corruption in democratic countries~\cite{dollar2001arewomen, merkle2023genderpart1, merkle2023genderpart2, barnes2024introduction}, suggesting that increased female representation in public offices and civil service also serves as a potential anti-corruption measure~\cite{swamy2001gender, karim2011madame, kahn2013mexican}. Research also indicates reductions in procurement-related corruption in French municipalities led by female mayors, although reelected female mayors exhibit no significant difference in corruption levels compared to their male counterparts~\cite{bauhr2021willwomen}. This finding suggests that, over time, females in leadership positions may integrate into existing political networks or establish new ones that are not necessarily less corrupt~\cite{esarey2018womens, bauhr2021willwomen}.

Despite numerous recent studies on the dynamics and inner workings of political corruption networks~\cite{ribeiro2018thedynamical, diviak2019structure, luna2020corruption, granados2021corruption}, organized crime~\cite{duijn2014relative, dacunha2018topology}, mafia groups~\cite{varese2013structure, cavallaro2020disrupting}, and other offenses committed through associations of individuals~\cite{dacunha2020assessing, elbahrawy2020collective, bouchard2020collaboration, chiang2024blend}, the role of gender in such networked crimes, particularly political corruption, remains relatively underexplored. A notable exception is the work by Smith~\cite{smith2020exogenous}, which examined gender disparities in Chicago's organized crime before and after the alcohol prohibition of 1920, revealing that the prohibition expanded opportunities for males while largely excluding females. This emerging research field holds significant potential for advancing academic understanding and equipping security forces with better tools for decision-making, particularly when enhanced by collaboration between law enforcement and academia. For instance, an empirical investigation into criminal networks assessed the impact of law enforcement interventions on the structure and functioning of a dark web pedophile network~\cite{dacunha2020assessing}, showing that the resulting dismantling resembled random node removal and was less effective than more optimized removal strategies. Other studies~\cite{lopes2022machine, ribeiro2023deep} showed that leveraging criminal network structures in combination with machine learning approaches yields impressive accuracy in recovering missing criminal partnerships, distinguishing types of criminal associations, predicting money flow among criminals, and even anticipating future partnerships and re-offending behavior. Recent studies analyzing security intelligence data~\cite{toledo2023multiplex, toledo2025outlier} proposed a novel protocol for network disruption through key-agent identification, showing its effectiveness in enabling law enforcement to identify and assess criminal activity. Another recent study on a dark web pedophile ring~\cite{divakarmurthy2024unravelling} used network tools to analyze user behavior, revealing crucial insights into temporal engagement patterns, content preferences, and user clusters. 

Research on networked crime has thus demonstrated that the structure of criminal networks is highly informative about various nuances of criminal activity~\cite{bao2025deterrence}. Therefore, it is critical to analyze possible disparities in the roles occupied by males and females within these networks. This issue parallels other gender-based analyses in fields such as education~\cite{colleen2018gender, bao2025can}, scientific careers~\cite{zeng2016differences, huang2020historical}, representation in literary works~\cite{stuhler2024thegender}, membership in artistic and scientific academies\cite{bao2022reform, card2023gender, bao2024gender}, and online biographies~\cite{wagner2021itsamans}, to name just a few, and may offer valuable perspectives on gender differences in networked crime. Here, we analyze data from political corruption scandals in Brazil~\cite{ribeiro2018thedynamical} and Spain~\cite{martins2022universality}, spanning from the 1990s to the 2010s, to construct corruption networks in which agents are connected when they collaborate in the same scandal. Gender labels are assigned to each network node based on the identification of individual names, allowing us to examine the structural roles of males and females within these networks. Consistent with the criminology literature, we find that males constitute the vast majority of corrupt agents, with females representing only 10\% of nodes in the Brazilian network and 20\% of nodes in the Spanish network -- fractions that are at least partly explained by very low participation in politics, high-ranking positions in the private sector, and other leadership roles~\cite{seo2017conceptual, reddy2019gender, piscopo2020, bao2022reform}. 

Based on this disparity, we adapt a model capable of reproducing several properties of these corruption networks~\cite{martins2022universality} to simulate networks where gender is randomly assigned to nodes while preserving the observed male-to-female ratios. These \textit{in silico} networks serve as null models for quantifying the significance of differences in network properties and determining whether these differences can be attributed solely to the disparity in gender representation. We examine the role of gender in centrality measures, network resilience, reoffending rates, and collaboration patterns, finding no significant gender differences overall, or that existing differences are at least partially explained by our null network model. The only exceptions refer to an overrepresentation of males among highly central individuals and a higher degree of gender homophily in collaboration patterns observed in the Spanish network, which cannot be accounted for by the null model. Our findings thus indicate a markedly low prevalence of females in political corruption cases; however, once this disparity is taken into account, the structural roles of these networked criminals appear to be largely invariant with respect to gender. Interestingly, these findings align with existing research on gender differences in scientific careers, where female and male faculties exhibit similar numbers of co-authors~\cite{zeng2016differences}, comparable annual productivity~\cite{huang2020historical}, and equivalent career-wise impact~\cite{huang2020historical}, once total publication numbers and career lengths are considered.

In what follows, we present these results in detail, beginning with a description of our data and the construction of the corruption networks, with an emphasis on the disparity in female participation. We then introduce our null network model, followed by an analysis of gender differences in centrality measures and network dismantling. Next, we explore gender differences in collaboration patterns within these networks. Finally, we conclude by contextualizing and summarizing our findings.

\section*{Results}\label{sec:results}

\subsection*{Data}

We begin by presenting the datasets used in our investigation, which encompass well-documented political corruption scandals in Brazil~\cite{ribeiro2018thedynamical} and Spain~\cite{martins2022universality}, spanning from the late 1980s to the late 2010s. These datasets were compiled from widely circulated magazines and newspapers in both countries and subsequently organized by the authors of Ref.~\cite{ribeiro2018thedynamical} for the Brazilian cases, and by a non-profit organization for the Spanish cases~\cite{casosaislados}. The data comprise the names of all individuals involved and the year of occurrence for 65 scandals in Brazil and 437 scandals in Spain. Using the list of involved individuals, we manually assign gender to each person based on their names, complemented by web searches to confirm the information. We then construct a complex network representation of these scandals where nodes represent individuals and edges connect pairs who partnered in at least one scandal. In both networks, nodes represent politicians, public servants, or private sector actors engaged in illegal practices that compromise the public good in return for monetary reward or the accrual of greater influence, access, or power. For the Spanish network, we further exclude isolated scandals involving a single individual. This process resulted in two corruption networks comprising 404 individuals in Brazil and 2695 in Spain. 

\subsection*{Gender prevalence}

\begin{figure*}[!t]
\centering
\includegraphics[width=0.95\linewidth]{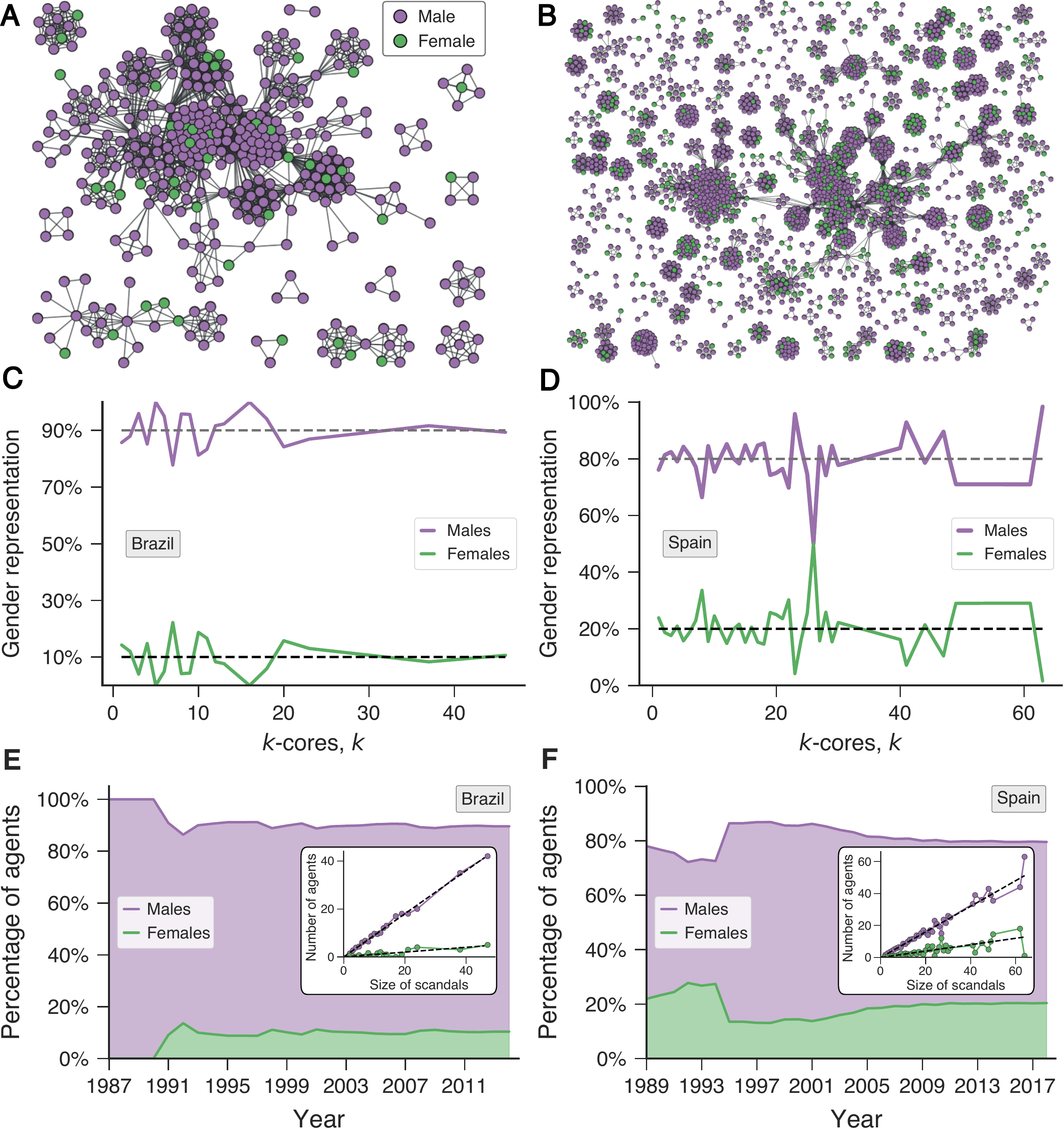}
\caption{Gender disparities in participation within corruption networks. Panels (A) and (B) present visualizations of corruption networks in Brazil and Spain. Nodes represent male (purple) and female (green) agents involved in corruption scandals, and connections among them indicate partnerships in at least one scandal. Females comprise 10\% of agents in Brazil and 20\% in Spain. Panels (C) and (D) show the proportions of males (purple line) and females (green line) across various $k$-core levels (defined as sets of nodes with at least $k$ connections) in the Brazilian and Spanish networks, respectively. In both panels, horizontal dashed lines indicate the average prevalence of males and females within each network. Panels (E) and (F) depict the evolution of the proportion of males (purple lines) and females (green lines) as the corruption networks in Brazil and Spain expand through the inclusion of new scandals. Insets show the average number of male (purple markers) and female (green markers) agents as a function of the scandal size (number of involved individuals). Dashed lines represent linear relationships in which the average number of agents is proportional to scandal size, with proportionality coefficients corresponding to the overall gender prevalence in each network.
}
\label{fig:1}
\end{figure*}

The Brazilian and Spanish networks are depicted in Figures~\ref{fig:1}A and \ref{fig:1}B, with nodes color-coded by gender (purple for males and green for females). These visualizations clearly show that female agents represent a small fraction of the nodes in both networks. Specifically, females account for $10\%$ and $20\%$ of the individuals in the Brazilian and Spanish networks, respectively. These proportions remain similar when considering only the largest connected component of the networks, where females constitute $9.4\%$ and $20.7\%$ of the Brazilian and Spanish networks, respectively. Moreover, Figures~\ref{fig:1}C and \ref{fig:1}D shows that the average prevalence of gender does not change across the different levels of connectivity obtained from the $k$-core decomposition~\cite{seidman1983network}, which progressively removes nodes with fewer than $k$ connections, creating a hierarchical structure of nodes, from the $1$-core (the entire set of nodes) to higher-order cores such as the $2$-core (nodes with at least two connections), and so on. We also examine the evolution of female participation as the networks expand over time due to the addition of new corruption cases. Figures~\ref{fig:1}E and \ref{fig:1}F show that, following a brief initial period with no females in the Brazilian network and a slightly higher fraction in the Spanish network, the proportion of females stabilizes and remains steady over time in both networks. Additionally, we analyze whether these proportions vary with the number of individuals involved in political scandals by estimating the average number of males and females as a function of scandal size. The insets of Figures~\ref{fig:1}E and \ref{fig:1}F show that the average numbers of male and female individuals scale proportionally with scandal size, with the proportionality rates corresponding to the overall gender incidence in the networks. These findings confirm that females are underrepresented in corruption networks and that this underrepresentation remains stable across both the hierarchical structures of connectivity and different scandal sizes.

\subsection*{Network centrality}

\begin{figure*}[!ht]
\centering
\includegraphics[width=1\linewidth]{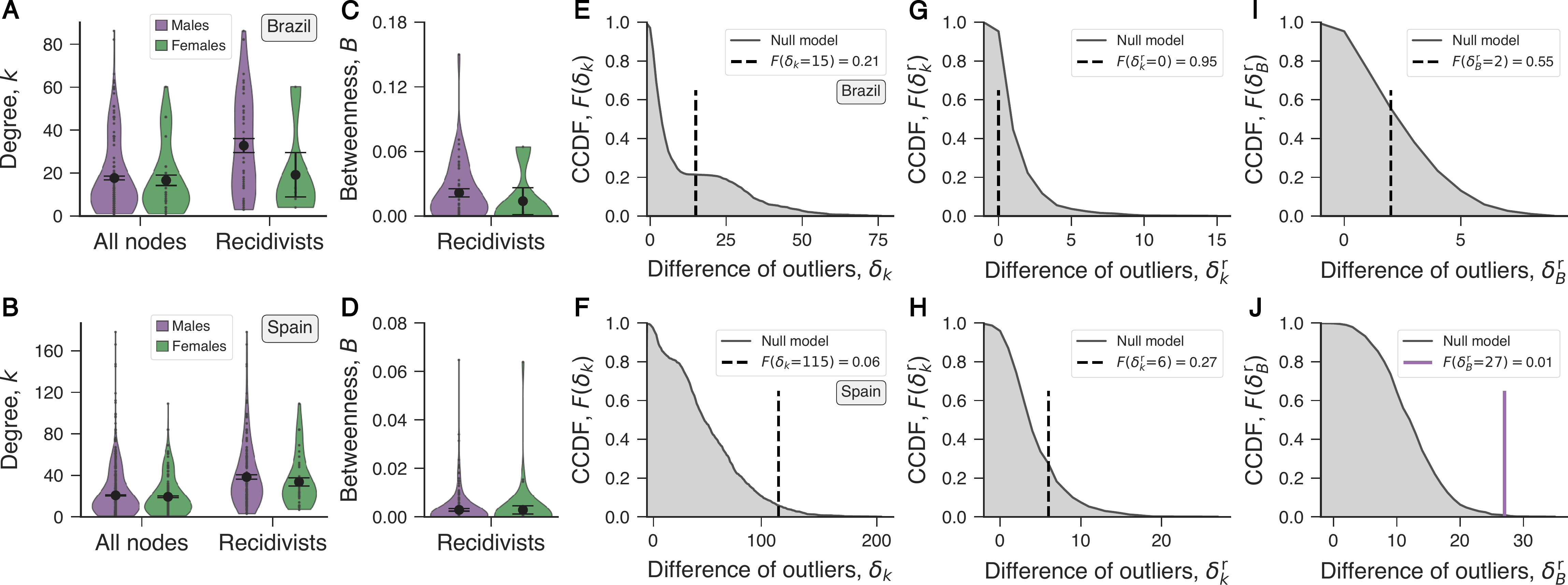}
\caption{Network centralities of males and females in corruption networks. Panels (A) and (B) show the degree centrality $k$ of males (purple) and females (green), with nodes further categorized into all agents and recidivist agents for the Brazilian and Spanish networks. Similarly, panels (C) and (D) present the betweenness centrality $B$ of recidivist agents. Error bars represent the average values of the centrality measures along with their standard errors of the mean. Violin plots illustrate the data distributions, with individual observations depicted as small dots. Panels (E)-(J) compare the difference between the number of males and females identified as outliers based on degree centrality ($\delta_k$ and $\delta_{k}^{\rm r}$ for recidivists) and betweenness centrality ($\delta_{B}^{\rm r}$) for the Brazilian (first row) and Spanish networks (second row). Vertical lines indicate the empirical differences observed in each network, while the filled curves correspond to the complementary cumulative distribution functions [CCDFs, $F(\delta)$, with $\delta\in(\delta_k, \delta_k^{\rm r}, \delta_B^{\rm r})$] of the same differences estimated from the null network model. Legends display the probability of observing the same or a large excess of male outliers in the null model.}
\label{fig:2}
\end{figure*}

In addition to examining gender prevalence within these networks, we assess whether males and females occupy distinct roles and functions by analyzing the degree ($k$) and betweenness ($B$) centrality measures~\cite{newman2018networks}. We estimate the average values of these measures by grouping nodes according to gender, further distinguishing whether individuals are recidivists. Recidivists, or re-offenders, defined as individuals involved in multiple corruption scandals, play a crucial role in shaping the structure and dynamics of corruption networks by linking separate scandals~\cite{martins2022universality}. Figures~\ref{fig:2}A-\ref{fig:2}D present the average values of both centrality measures (circle markers with error bars) alongside their distributions (shown as violin plots) separated by gender (purple for males and green for females). On average, male individuals have $18\pm1$ partners (with $\pm$ denoting the standard error of the mean), while females show a slightly lower average of $17\pm2$ partners in the Brazilian network. A similar trend is observed in the Spanish network, where males average $21\pm1$ partners, compared to $19\pm1$ for females. Among recidivists, males also exhibit a higher average number of partners than females in both networks, with values of $33\pm3$ vs. $19\pm10$ and $38\pm2$ vs. $34\pm4$ for recidivist males and females in the Brazilian and Spanish networks, respectively. For betweenness centrality, we calculate the average values only for recidivists, as non-recidivists have null betweenness given they do not intermediate any shortest paths. The average betweenness of female recidivists [$(1.39\pm1.26)\times10^{-2}$ and $(2.87\pm1.65)\times10^{-3}$] is also slightly smaller than that of males [$(2.16\pm0.38)\times10^{-2}$ and $(2.87\pm0.50)\times10^{-3}$] in both networks (Brazilian and Spanish).

The analysis of centrality measures suggests that, on average, males not only have more partners but also intermediate more shortest paths within these networks. To systematically test these hypotheses, we employ bootstrap tests for the equality of means~\cite{efron1994introduction}, finding that none of the observed gender differences in the average values of either centrality measure are statistically significant at the 95\% confidence level (see Supplementary Table~S1 for all $p$-values). Additionally, Mann-Whitney tests for equality in distribution~\cite{conover1999practical} do not allow rejection of the null hypothesis that the distributions of these centrality measures are equal between the genders at the same confidence level (see Supplementary Table~S2 for all $p$-values).

To further corroborate these results, we adapt a model capable of replicating several properties of corruption networks~\cite{martins2022universality} to generate synthetic networks where gender is randomly assigned to nodes, maintaining the observed proportions of males and females in each network. As detailed in Methods, the model begins with an empty network and incrementally adds fully connected graphs, simulating the emergence of new scandals. Nodes within these subgraphs have a probability of being assigned as re-offenders (the recidivism rate), and when a re-offender appears, her or his new partners are connected together to that re-offender in the network. This process continues until the number of scandals in the synthetic networks matches those observed in the empirical networks. Additionally, following~\cite{martins2022universality}, we fine-tune the model parameters to best replicate the properties of each empirical network. Once the synthetic networks are generated, gender is randomly assigned to each node according to the empirical proportions of males and females (10\% in the Brazilian and 20\% in the Spanish networks). This method thus generates a null model in which gender is not associated with any structural property of the corruption networks, reflecting only the empirical disparity in gender representation. Indeed, simulations of this model mimic the empirical behavior and produce networks where degree and betweenness centrality measures display the same average values and distributions for both genders (Supplementary Figure~S1). 

Yet, the results of Figures~\ref{fig:2}A-\ref{fig:2}D indicate that the centrality distributions are highly skewed, with certain individuals exhibiting centrality values substantially higher than the average. This prompts the question of whether male and female agents are equally represented among these highly central individuals. To investigate this possibility, we calculate the difference between the numbers of males and females identified as outliers based on degree ($\delta_k$ and $\delta_k^{\rm r}$ for recidivists) and betweenness ($\delta_B^{\rm r}$, only for recidivists) centralities. Outliers are defined using Tukey's fences rule~\cite{tukey1977exploratory}, where any observation exceeding the third quartile by more than $1.5$ times the interquartile range is classified as an outlier. In the Brazilian network, we identify an excess of $15$ males among degree outliers ($\delta_k=15$), no difference among recidivists outliers in degree ($\delta_k^{\rm r}=0$), and an excess of $2$ males among outliers in betweenness centrality ($\delta_B^{\rm r}=2$). In the Spanish network, there is an excess of $115$ males among degree outliers ($\delta_k=115$), an excess of $6$ males among recidivist outliers in degree ($\delta_k^{\rm r}=6$), and an excess of $27$ males among betweenness outliers ($\delta_B^{\rm r}=27$). These results suggest that males constitute the majority of highly central individuals; however, part of this male overrepresentation may be due to the significantly higher proportion of males in these corruption networks.

To test this hypothesis, we generate one thousand simulated networks from our null model for each country, evaluating the differences in the numbers of males and females identified as outliers. We then calculate the complementary cumulative probability distributions of these differences [$F(\delta)$, where $\delta\in(\delta_k, \delta_k^{\rm r}, \delta_B^{\rm r})$], as shown in Figures~\ref{fig:2}E-\ref{fig:2}J. Most of the distribution masses lie in the positive range, confirming that the higher prevalence of males is indeed partially responsible for the emergence of more male outliers. These distributions further allow us to quantify how rare the empirical differences between the number of male and female outliers are, analogously to a $p$-value in hypothesis tests. In the Brazilian network, the empirical differences (represented as vertical lines in Figures~\ref{fig:2}E, \ref{fig:2}G, and \ref{fig:2}I) all have higher probabilities of randomly occurring in our null model, suggesting that they can be fully explained by the higher proportion of males in the network. Similarly, in the Spanish network, the excess of male outliers in degree for recidivists has a $27\%$ probability of occurring randomly (vertical line in Figure~\ref{fig:2}H). However, for degree and betweenness centrality outliers (vertical lines in Figures~\ref{fig:2}F and \ref{fig:2}J), the probabilities of observing the empirical differences in our null models are significantly lower. Specifically, the probability of observing $115$ or more male degree outliers is estimated at $6\%$, and the probability of finding $27$ or more male betweenness outliers is $1\%$. Although the $6\%$ probability for degree outliers falls within the $95\%$ confidence level, these findings support, at least partially, the hypothesis that males are overrepresented among highly connected individuals in the Spanish network, even beyond what numerical differences in gender prevalence would suggest. We also note that the Brazilian network is significantly smaller than the Spanish network, which may explain the lack of significance as random fluctuations become more pronounced in smaller systems.

\subsection*{Network resilience}

\begin{figure*}[!t]
\centering
\includegraphics[width=1\linewidth]{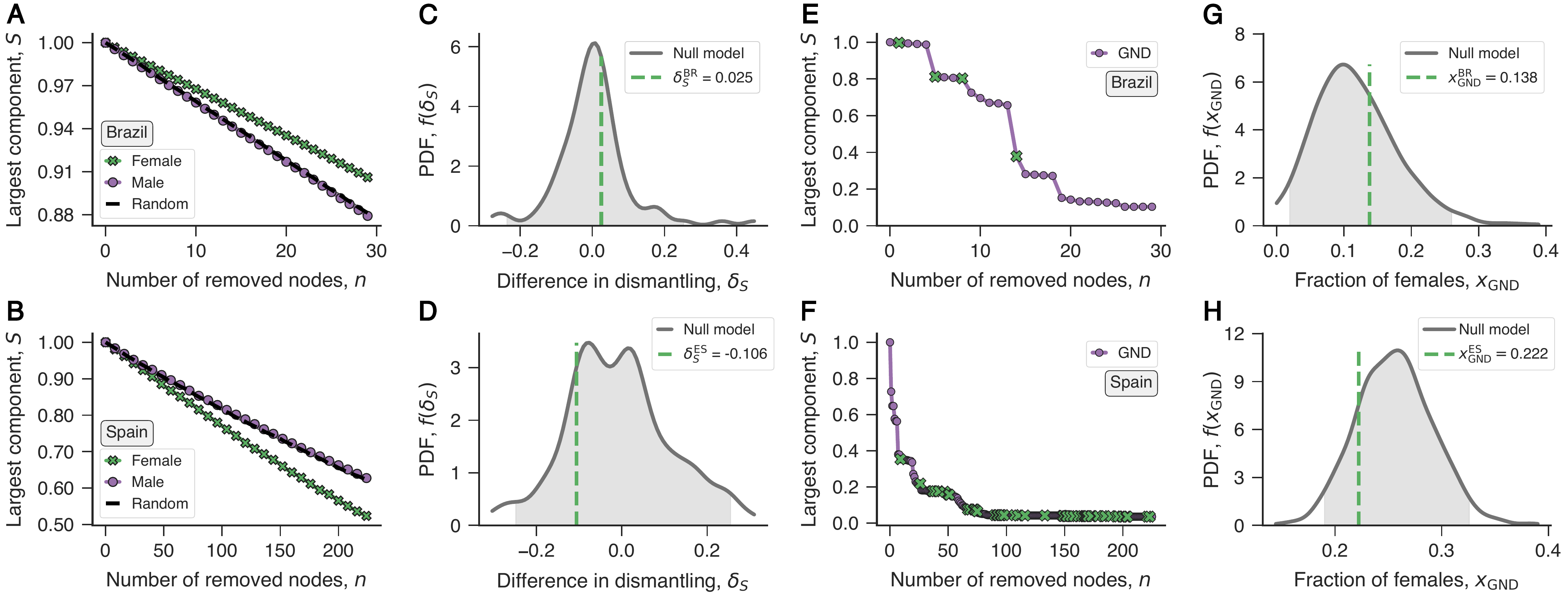}
\caption{Network resilience under the removal of male and female agents. Fraction of the largest connected component $S$ of the (A) Brazilian and (B) Spanish networks as a function of the number of removed nodes $n$ following three random dismantling strategies: randomly removing all females (green crosses), randomly removing the same number of males (purple circles), and randomly removing the same number of agents regardless of gender (dashed lines). The curves represent the average values of $S$ calculated from $1{,}000$ independent realizations of each dismantling strategy. Probability distribution functions [PDFs, $f(\delta_S)$] comparing the differences $\delta_S$ in the final fraction of the largest component after randomly removing females and randomly removing nodes regardless of gender in $1{,}000$ simulations of networks generated from our null model for the (C) Brazilian and (D) Spanish networks. Fraction of the largest connected component $S$ of the (E) Brazilian and (F) Spanish networks as a function of the number of removed nodes $n$ following the generalized network dismantling (GND) algorithm proposed by Ren~\textit{et al.}~\cite{ren2019generalized}. Green crosses indicate the removal of female agents, and purple circles represent the removal of male agents. The GND algorithm is deterministic with nodes sequentially removed (up to the total number of females initially present in the giant component of each network) in an optimal order designed to cause maximum disruption. Probability distribution functions [PDFs, $f(x_{\text{GND}})$] of observing a fraction $x_{\text{GND}}$ of female agents among the nodes selected for removal by the GND algorithm across $1{,}000$ simulations of our null model for the (G) Brazilian and (H) Spanish corruption networks. The empirical fractions are depicted by vertical lines, with shaded regions corresponding to 95\% confidence intervals estimated from the null model. 
}
\label{fig:3}
\end{figure*}

We also examine the potential influence of gender on the resilience of the Brazilian and Spanish corruption networks by performing random dismantling and targeted attacks on their largest components. For random dismantling, we consider three distinct strategies: randomly removing all females, randomly removing the same number of males, and randomly removing the same number of agents regardless of gender. Each strategy is replicated $1{,}000$ times per network, and the attack's impact is assessed by calculating the average fraction of the largest connected component $S$ as a function of the number of removed nodes $n$, as depicted in Figure~\ref{fig:3}A for the Brazilian network and Figure~\ref{fig:3}B for the Spanish network. In both cases, the average behavior of $S$ when removing only males (purple circles) or when removing nodes regardless of gender is similar due to the significantly higher prevalence of males. Interestingly, the random removal of females produces different results between the two networks. In the Brazilian network, removing all females causes $3\%$ less damage to the giant component than removing the same number of males or nodes regardless of gender. In contrast, in the Spanish network, removing all females causes $16\%$ more damage to the giant component than the two other strategies. To verify the significance of these findings, we compare the effects of randomly removing females with those of randomly removing nodes regardless of gender in networks generated from our null model, calculating the differences in the final fraction of the largest component ($\delta_S$), and their distributions across $1{,}000$ simulations. Figures~\ref{fig:3}C and \ref{fig:3}D show that observing $3\%$ less damage to the giant component when removing only females in the giant component of the Brazilian network or the $16\%$ more damage caused by the removal of all females in the giant component of the Spanish network falls within the 95\% confidence intervals derived from our null model. Therefore, there is no significant difference between randomly removing only females and only males to the resilience of both corruption networks. The same conclusions are obtained when evaluating the attack damages by calculating the average value of the inverse of the shortest path lengths in the giant components of these networks (Supplementary Figure~S2).

In addition to random dismantling, we further consider targeted attacks using the deterministic dismantling algorithm proposed by Ren~\textit{et al.}~\cite{ren2019generalized}, addressing the generalized network dismantling (GND) problem. This approach aims to identify an optimal set of nodes whose removal (subject to cost constraints) reduces the largest connected component to a specified maximum size. Since the algorithm of Ren \textit{et al.} relies on a recurrent spectral partitioning approach, it offers complementary insights into the structural roles of males and females that may not be captured by the random dismantling and centrality analysis. Figures~\ref{fig:3}E and \ref{fig:3}F depict the fraction of the largest components of the Brazilian and Spanish networks as nodes selected by the algorithm are sequentially removed (up to the total number of females initially present) in an optimal order designed to cause maximum disruption. In these figures, the removal of male nodes is indicated by purple circles, while the removal of female nodes is shown by green crosses. As expected, these targeted attacks inflict significantly greater damage on the giant components of both networks compared to random dismantling strategies. Regarding gender differences, in the Brazilian network, $4$ females out of the $29$ initially present are removed, corresponding to $13.8\%$, a fraction slightly higher than the overall female prevalence in the giant component ($9.4\%$). Similarly, in the Spanish network, the GND algorithm selects 50 females for removal out of the 225 initially present, yielding a fraction slightly higher than the overall female prevalence in the giant component ($22.2\%$ vs. $20.7\%$). Once again, we use our null model to assess the significance of these differences by estimating the probability distributions of observing a fraction $x_{\text{GND}}$ of females among the nodes selected for removal across $1{,}000$ simulations for each corruption network. These distributions are shown in Figures~\ref{fig:3}G and \ref{fig:3}H, indicating that both empirical differences are fully accounted for by our null model. The same results are obtained when considering targeted network attacks based on degree and betweenness centralities, as well as when quantifying damage using the average value of the inverse of the shortest path lengths (Supplementary Figure~S3). Therefore, as with random dismantling, gender does not play a significant role in targeted network attacks.

\subsection*{Collaboration patterns}

\begin{figure*}[!t]
\centering
\includegraphics[width=1\linewidth]{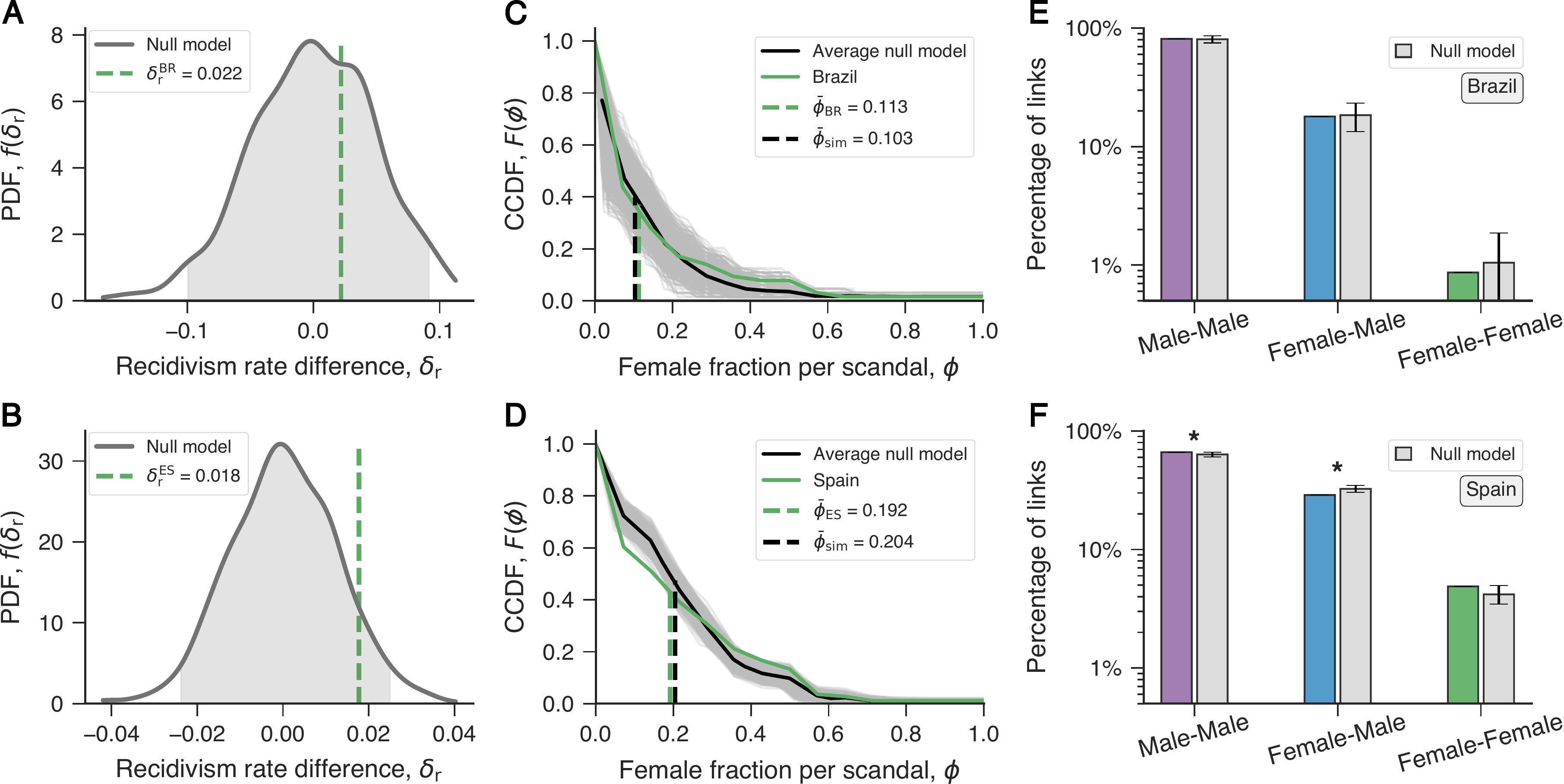}
\caption{Collaboration patterns of males and females within corruption networks. Probability distribution functions [PDFs,~$f(\delta_r)$] of differences in recidivism rates $\delta_r$ between males and females estimated from simulation of our null model with parameters tailored to the (A) Brazilian and (B) Spanish data. Vertical lines indicate the empirical differences calculated for the Brazilian ($\delta_r^{\text{BR}}=0.022$) and Spanish ($\delta_r^{\text{ES}}=0.018$) networks. The shaded regions represent $95\%$ confidence intervals estimated from the simulated data for each country. Complementary cumulative distribution functions [CCDFs, $F(\phi)$] of the fraction $\phi$ of females involved in corruption scandals in the (C) Brazilian and (D) Spanish data. Green curves represent the empirical distributions, gray curves show the distributions estimated from each of the $1{,}000$ simulations of the null model for each country, and the black curves correspond to the average behavior of the simulations. Green vertical lines indicate the empirical average fractions ($\bar{\phi}_{\rm BR} = 0.11$ for Brazilian and $\bar{\phi}_{\rm ES} = 0.19$ for Spanish scandals), while black vertical lines show the corresponding averages calculated from all simulations. Percentage of male-male, female-male, and female-female links in the (E) Brazilian and (F) Spanish corruption networks on a logarithmic scale. Colored bars indicate the empirical percentages, while gray bars show the averages obtained from $1{,}000$ simulations of the null model for each country. Error bars represent the 95\% confidence intervals calculated from the null model simulations, with asterisks above the bars indicating statistically significant differences.}
\label{fig:4}
\end{figure*}

In our final analysis, we examine collaboration patterns of both genders in corruption networks. Results so far indicate that females and males display a similar number of collaborators, including among recidivists. However, males are overrepresented among highly central individuals in the Spanish network. This prompts the question of whether there are differences in the recidivism rates (defined here as the proportion of individuals involved in multiple corruption cases) between males and females. In the Brazilian network, $14.1\%$ of males and $11.9\%$ of females participate in more than one corruption scandal, while in the Spanish network, $9.04\%$ of males and $7.27\%$ of females are recidivists. Thus, the recidivism rate of males is $\approx2\%$ higher than that of females in both networks. Once again, we assess the significance of these differences by simulating $1{,}000$ networks using our null model with parameters tailored to the data of each country. We then calculate the differences $\delta_r$ in the recidivism rate between genders and estimate their probability distributions. Figures~\ref{fig:4}A and \ref{fig:4}B display these distributions for Brazilian and Spanish simulated networks, where shaded regions represent $95\%$ confidence intervals for $\delta_r$, and the vertical lines indicate the empirical differences. As the empirical differences fall within the confidence intervals, they are not statistically significant, suggesting no clear evidence that males have higher recidivism rates than females. 

We also explore the cumulative distributions $F(\phi)$ of the fractions $\phi$ of females involved in each corruption scandal, as depicted in Figures~\ref{fig:4}C for Brazilian scandals and \ref{fig:4}D for the Spanish scandals (green lines). These figures show that scandals predominantly involving females ($\phi>0.5$) are rare ($3\%$ and $4\%$ in Brazilian and Spanish data, respectively); however, the average values of $\phi$ ($\bar{\phi}_{\rm BR} = 0.11$ for Brazilian and $\bar{\phi}_{\rm ES} = 0.19$ for Spanish scandals) closely align with the overall proportions of females in both networks ($0.10$ and $0.20$, respectively). Using the Mann-Whitney test~\cite{conover1999practical}, we compare these empirical distributions with those derived from $1{,}000$ simulations of our null model (gray curves in Figures~\ref{fig:4}C and \ref{fig:4}D). The test fails to reject the null hypothesis of equality between the simulated and empirical distributions for the Brazilian data in 98\% of the comparisons ($p$-value $<0.05$ in only $2\%$ of the comparisons, Supplementary Figure~S4), and in the Spanish data, the hypothesis is not rejected in 61\% of the comparisons ($p$-value $<0.05$ in $39\%$ of the comparisons, Supplementary Figure~S4). Furthermore, the average values of $\phi$ and $F(\phi)$ across all simulations closely match their empirical counterparts, with only small systematic deviations observed in the Spanish scandals. While the prevalence of females in Spanish scandals appears marginally higher than that predicted by the null model, the fraction of scandals predominantly involving females ($\phi>0.5$) is not significantly different from the simulated results (Supplementary Figure~S5). These findings confirm that scandals predominantly involving females are rare, and this underrepresentation can be fully explained by their overall lower representation in the networks.

Still analyzing collaboration patterns, we calculate the percentage of links exclusively between males, between females and males, and exclusively between females to quantify gender homophily within the corruption networks. As before, we use $1{,}000$ simulated replicas of the Brazilian and Spanish networks generated from our null model as baselines for comparisons with empirical values. The colored bars in Figures~\ref{fig:4}E and \ref{fig:4}F show the empirical proportions of each link type, while the grey bars represent the null model simulations, with error bars indicating $95\%$ confidence intervals. In the Brazilian network, the proportions of male-male, female-male, and female-female links are $81.2\%$, $17.9\%$, and $0.9\%$, respectively. These empirical percentages closely align with the average values obtained from the null model and do not significantly differ from them (Supplementary Figures~S6A-S6C), suggesting that the observed degree of homophily in the Brazilian network is fully explained by the null model. In contrast, in the Spanish network, $66.3\%$ of links are exclusively between males, $28.8\%$ connect females and males, and $4.9\%$ are exclusively between females. These percentages deviate more substantially from the null model averages of $63.3\%$, $32.5\%$, and $4.2\%$ for male-male, female-male, and female-female connections, respectively. The surplus of male-male links and the deficit of female-male connections are statistically significant, as the empirical values lie outside the $95\%$ confidence intervals (Supplementary Figures~S6D and S6E). In turn, although the excess of female-female links is not strictly statistically significant, the empirical value ($4.9\%$) closely approaches the upper limit of the $95\%$ confidence interval ($3.4\%$ to $4.9\%$, Supplementary Figures~S6F). Thus, our data provide evidence for the existence of a higher degree of gender homophily in the Spanish network that cannot be fully accounted for by the null model.

\section*{Discussion and conclusions}\label{sec:conclusion}

Corruption, broadly defined as ``the abuse of entrusted power for private benefits or gains,''~\cite{luna2020corruption} is a pervasive criminal behavior that remains highly resistant to control~\cite{mungiu2017time, unodc2023annual}. The vast sums of money involved in corruption cases, alongside the high costs of efforts to combat this crime, have driven much of the literature on corruption and organized crime toward economic and sociological analyses~\cite{mungiu2017time}. In parallel, the rise of network science~\cite{vespignani2018twenty, newman2018networks} has recently fostered a growing body of research exploring the dynamics and structure of networked crimes~\cite{duijn2014relative, ribeiro2018thedynamical, dacunha2018topology, luna2020corruption, elbahrawy2020collective, dacunha2020assessing, granados2021corruption, lopes2022machine, ribeiro2023deep, toledo2023multiplex, divakarmurthy2024unravelling}. Our study contributes to this growing field by addressing the underexplored role of gender in corruption networks, thus complementing the extensive literature documenting the significantly higher offending rates of males across various crime types~\cite{dollar2001arewomen, swamy2001gender, bennett2005explaining, kruttschnitt2013gender, rivas2013anexperiment}, as well as the suggested lower tolerance of females for dishonest and corrupt behavior~\cite{swamy2001gender, dollar2001arewomen, karim2011madame, chaudhuri2012gender, rivas2013anexperiment, kahn2013mexican, bauhr2021willwomen}.

Our findings align with the criminology literature, demonstrating that males constitute the vast majority of individuals in political corruption networks, with females representing only $10\%$ of nodes in the Brazilian network and $20\%$ of nodes in the Spanish network. Moreover, we observed that the proportion of females involved in corruption scandals remained stable over time, across hierarchical connectivity structures, and irrespective of scandal size. Given that most individuals involved in corruption scandals are politicians and high-ranking employees in both public and private sectors, the underrepresentation of female agents may, at least in part, be attributable to their comparatively low participation in politics, high-ranking positions, and other leadership roles~\cite{seo2017conceptual, reddy2019gender, piscopo2020, bao2022reform}. Indeed, during the same period covered by our data, females constituted, on average, $10.6\%$ and $26.4\%$ of parliamentarians in the upper and lower chambers in Brazil and Spain~\cite{cardoso2020gender, ipu2024women}, respectively -- proportions that parallel the female participation in our corruption networks. However, corruption network participants are not exclusively political representatives and the complete networks capturing all relevant interactions between politics and other sectors remain inaccessible, leaving the associated gender distributions unknown. Moreover, politicians are not ordinary individuals~\cite{piscopo2020}, and political corruption represents a highly specialized form of criminality. Thus, selection bias associated with the occupation of political positions, as well as bias related to the emergence of individuals involved in illicit activities, may also influence the observed gender differences in corruption network participation. Accordingly, it remains a challenge for future studies to determine whether disparities in political participation and leadership roles fully account for the underrepresentation observed.

Interestingly, the level of female participation in these corruption networks is somewhat comparable to that observed in Chicago organized crime prior to alcohol prohibition in 1920, where females constituted 18\% of networked individuals~\cite{smith2020exogenous}. However, following prohibition, female representation dropped to just 4\%, despite the network nearly tripling in size~\cite{smith2020exogenous}. A similarly low level of female participation was observed in organized crime groups involved in serious offenses in a region of England, where females accounted for 6\% of individuals~\cite{campana2022determinants}. In the case of Chicago, Smith~\cite{smith2020exogenous} argued that the exogenous shock of alcohol prohibition increased profits, violence, and risk, leading to new organizational structures that concentrated power and opportunities among males, largely excluding females from criminal activities. While our findings do not indicate significant variations in female involvement across both corruption networks, the potential impact of similar exogenous shocks -- such as anti-corruption measures or large-scale police operations -- on these networks warrants further investigation in future research.

To explore how this gender disparity translates into differences in network properties of males and females, we considered a model capable of replicating several aspects of corruption networks, generating simulated networks that reflect the observed gender imbalance but without tying gender to any specific network property. These simulations thus served as null models, providing baselines for statistical comparisons and revealing differences that cannot be attributed solely to the disparity in gender representation. Despite the substantial imbalance in gender representation, our analysis found that females and males exhibit similar averages and distributions in terms of degree and betweenness centralities. However, the number of males emerging as outliers for both centrality measures was significantly greater than that of females, with these differences not being fully explained by the female underrepresentation in our null models, particularly in the Spanish network. The lower occurrence of females among centrality outliers also aligns with the absence of females in the criminal elite during the alcohol prohibition era in Chicago~\cite{smith2020exogenous}. Gender also did not significantly affect the resilience of corruption networks, whether through random dismantling or targeted attacks on their largest components. Specifically, we found that randomly removing only females, only males, or removing agents irrespective of gender caused similar levels of damage to these networks. Similarly, the fraction of females selected for removal by a deterministic algorithm addressing the generalized network dismantling problem~\cite{ren2019generalized} was indistinguishable from the values obtained in the null model simulations. 

Regarding collaboration patterns, our findings indicated that males are more likely than females to be involved in multiple scandals and that the proportion of scandals predominantly formed by females is small in both networks. However, these differences are fully explained by the overall lower female representation, as captured by the null network model. Although our simulations fully explain the gap in recidivism rates, it is noteworthy that female politicians are often subjected to higher levels of voter accountability -- given the general perception of females as more honest than males~\cite{rivas2013anexperiment, chaudhuri2012gender} -- which may further disincentivize their involvement in multiple scandals~\cite{esarey2018womens, eggers2018corruption}. We further investigated gender homophily within the corruption networks by estimating the percentage of links exclusively between males, between females and males, and exclusively between females, both from the empirical networks and their null model simulations. In both networks, male-male connections accounted for the vast majority of network connections, followed by female-male connections, while exclusively female connections were the least common. In the Brazilian network, the degree of homophily quantified by the prevalence of these three connection types was fully explained by the null model. Conversely, in the Spanish network, male-male connections exceeded and male-female connections fell below null model expectations, indicating that the degree of gender homophily is not solely explained by female underrepresentation. This pattern aligns with previous research on organized crime, where homophilic tendencies have been observed in relation to ethnicity~\cite{grund2012ethnic}, age~\cite{warr1996organization, campana2022determinants}, and gender~\cite{warr1996organization, smith2020exogenous, campana2022determinants}. Specifically, Campana and Verese identified gender as a determinant of membership in criminal groups, with individuals of the same gender more likely to belong to the same group~\cite{campana2022determinants}. Similarly, Smith~\cite{smith2020exogenous} observed mild assortative mixing by gender, where females tend to connect with males, while males predominantly connect with other males in the context of organized crime in Chicago during the alcohol prohibition.

Taken together, our findings indicate that the participation of female agents in political corruption is markedly low; however, once this disparity is accounted for, the structural roles of these networked criminals appear largely invariant with respect to gender, with notable exceptions concerning the presence of highly central individuals and the degree of homophily in criminal associations. Although not a perfect analogy, these findings align with existing research on gender differences in scientific careers, where female and male faculties exhibit similar numbers of co-authors~\cite{zeng2016differences}, comparable annual productivity~\cite{huang2020historical}, and equivalent career-wise impact~\cite{huang2020historical}, once total publication numbers and career lengths are considered. 

Our research is not without limitations. A primary challenge lies in the difficulty of acquiring reliable data on corruption networks and organized crime, due to legal confidentiality or the covert nature of criminal activities. As with many datasets of this type, corruption data remains inherently incomplete, as it is challenging to ensure that all individuals involved in corruption scandals are identified through investigations. The temporal span of our networks (from the late 1980s to the late 2010s) may also influence our results. Future research using more recent data, in light of the rapid developments in female representation and evolving societal norms regarding gender equality, may eventually reveal emerging trends despite the observed stability in gender prevalence. Furthermore, the limited size of our corruption networks may have constrained the detection of statistically significant differences, particularly in the Brazilian context, where random fluctuations in the null model were more pronounced. Moreover, our conclusions are based on political corruption within two Western nations, and although these countries differ in political systems and development levels, future research should expand to other regions to assess the generalizability of our findings. Another limitation is the exclusion of nonbinary identities from our analysis; although such data are undoubtedly challenging to collect, future work should strive to incorporate these identities for a more comprehensive understanding of gender roles in corruption networks. Additional studies could also explore the temporal dynamics of centrality and collaboration within corruption networks, considering their growth and evolution as well as the role of gender in the emergence of isolated scandals and individuals. Equally important is the examination of other networked crimes, such as money laundering, illicit trade, and cybercrime. Despite these constraints, we believe this work improves our understanding of gender disparities and commonalities in the roles individuals occupy within corruption networks.

\section*{Methods}\label{sec:methods}

We adapt the corruption network model proposed in Ref.~\cite{martins2022universality} to generate a null model in which gender labels are randomly assigned based on the empirical gender proportions. The original model was designed to replicate several characteristics observed in corruption networks, including exponential degree distributions, high clustering coefficients, and the small-world property. The model begins with an empty synthetic network that grows by iteratively adding complete graphs representing corruption scandals. To mimic empirical behavior, the size $s$ (number of people involved) in these scandals is drawn from the exponential distribution, $P(s) = \frac{1}{s_c} e^{-s/s_c}$, where $s_c$ is a parameter defining the typical size of corruption scandals. Additionally, to further replicate empirical patterns, the model assumes that the number of recidivist agents ($R$) increases linearly with the total number of nodes ($N$) according to $R = \alpha N + \beta$, where $\alpha$ controls the recidivism rate and $\beta$ determines the minimal number of individuals in the network required to observe the first recidivists. Each time new recidivists are introduced, the model randomly selects individuals from the network to become recidivists and incorporates them into the next scandal added to the network. Furthermore, individuals who are already recidivists are selected with a small probability $p_{rr}$. This process continues until the number of corruption scandals in the simulated networks matches that observed in the empirical networks. Once the network is complete, genders are randomly assigned to all nodes based on a Bernoulli distribution, where the probability of a node being female is $p_f$.

Following Ref.~\cite{martins2022universality}, we set the model parameters to best reproduce the empirical properties of the corruption networks. For the Brazilian network, we use $\alpha=0.14$, $\beta=-11.5$, $s_c=7.51$, and $p_f=0.1$. For the Spanish network, we set $\alpha=0.09$, $\beta=-3.47$, $s_c=7.33$, and $p_f=0.2$. We further consider $p_{rr}=0.025$ for both networks.

\section*{Acknowledgements}

We acknowledge the support of the Coordena\c{c}\~ao de Aperfei\c{c}oamento de Pessoal de N\'ivel Superior (CAPES -- PROCAD-SPCF Grant 88881.516220/2020-01), the Conselho Nacional de Desenvolvimento Cient\'ifico e Tecnol\'ogico (CNPq -- Grant 303533/2021-8), and Slovenian Research and Innovation Agency (Javna agencija za znanstvenoraziskovalno in inovacijsko dejavnost Republike Slovenije) (Grants P1-0403 and N1-0232).

\section*{Author contributions statement}
A.A.B.P., A.F.M., M.V.P., S.G., C.M., M.P., and H.V.R. designed research, performed research, analyzed data, and wrote the paper.
 
\section*{Data availability}
The data used in the current study are available from Refs.~\cite{ribeiro2018thedynamical, martins2022universality} and the corresponding author upon reasonable request.

\bibliography{references}

\clearpage
\includepdf[pages=1-7,pagecommand={\thispagestyle{empty}}]{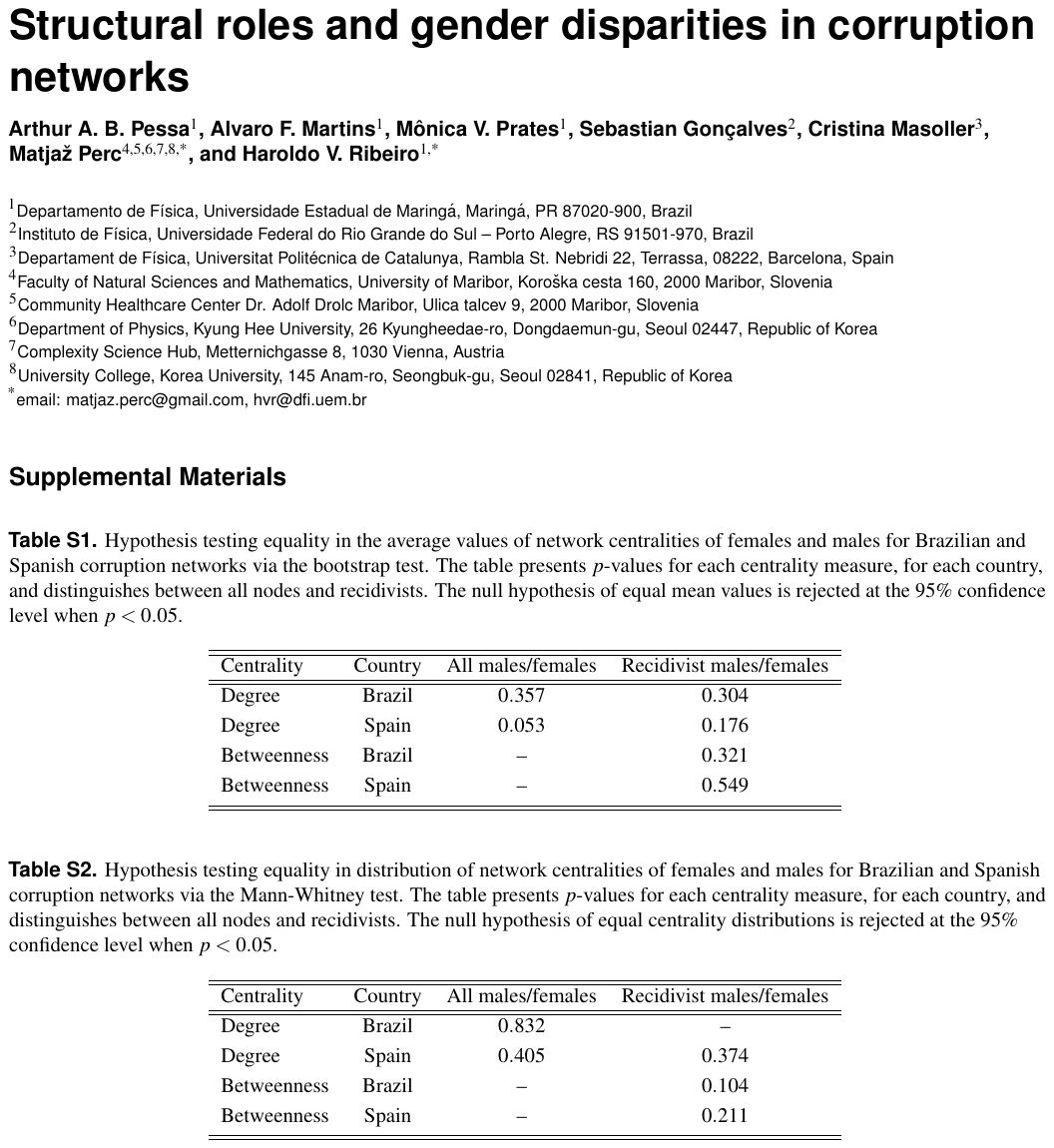}

\end{document}